\documentclass{article}
\usepackage[utf8]{inputenc}
\usepackage[T1]{fontenc}  
\usepackage{footmisc}
\usepackage{hyperref}     
\usepackage{url}          
\usepackage{booktabs}     
\usepackage{amsfonts}     
\usepackage{nicefrac}     
\usepackage{microtype}    
\usepackage{lipsum}
\usepackage{fancyhdr}     
\usepackage{graphicx}     
\graphicspath{{media/}}   

\usepackage{amsmath}
\usepackage{caption}
\usepackage{subcaption}

\usepackage{authblk}
\title{Sectional curvatures distribution of complexity geometry}

\author[1,2]{Qi-Feng Wu\thanks{wuqifeng199809@gmail.com}}

\affil[1]{Department of Physics and Center for Field Theory and Particle Physics,

Fudan University, Shanghai, 200433, China}
\affil[2]{Shenzhen Institute for Quantum Science and Engineering,

Southern University of Science and Technology, Shenzhen, 518055, China}

\begin{document}

\maketitle

\begin{abstract}
In the geometric approach to defining complexity, operator complexity is defined as the distance in the operator space. In this paper, based on the analogy with the circuit complexity, the operator size is adopted as the metric of the operator space where the path length is the complexity. The typical sectional curvatures of this complexity geometry are positive. It is further proved that the typical sectional curvatures are always positive if the metric is an arbitrary function of operator size, while complexity geometry is usually expected to be defined on negatively curved manifolds. By analyzing the sectional curvatures distribution for the $N$-qubit system, it is shown that surfaces generated by Hamiltonians of size smaller than the typical size can have negative curvatures. In the large $N$ limit, the form of complexity metric is uniquely constrained up to constant corrections if we require sectional curvatures are of order $1/N^2$. With the knowledge of states, the operator size should be modified due to the redundant action of operators, and thus is generalized to be state-dependent. Then we use this state-dependent operator size as the metric of the Hilbert space to define state complexity. It can also be shown that in the Hilbert space, 2-surfaces generated by operators of size much smaller than the typical size acting on typical states also have negative curvatures.

\end{abstract}

\tableofcontents

\section{Introduction}

Quantum complexity, as a concept that originated from quantum computation, plays an important role in the study of quantum gravity in recent years. It measures the number of simple quantum gates in a quantum circuit. Complexity analysis is a useful way to understand the firewall puzzle~\cite{Almheiri:2012rt}, which is known as the Harlow-Hayden conjecture~\cite{Harlow:2013tf}. As a boundary observable, it's conjectured that the gravitational dual of complexity is the volume or the action in the bulk, and can be used to probe the interior of black holes~\cite{Susskind:2014rva,Brown:2015lvg}. For other applications of complexity to understand quantum gravity, see e.g.~\cite{Susskind:2018tei,Bernamonti:2019zyy,Czech:2017ryf,Susskind:2019ddc,Brown:2019rox}.

Although holographic complexity has been extensively studied, it's still an open problem to define complexity appropriately. One of the main approaches to defining complexity is Nielsen's geometric approach~\cite{nielsen2005geometric,2007quant.ph..1004D}, called complexity geometry. In the geometric way to define complexity, the complexity is defined as the path length on the unitary group or Hilbert space, so a proper distance function is required. Various distance functions are proposed~\cite{2007quant.ph..1004D,Jefferson:2017sdb,Chapman:2017rqy,Brown:2017jil,Khan:2018rzm,Flory:2020eot,Flory:2020dja}, but the physical meaning of these distance functions is not clear enough. For other recent works, see\cite{Basteiro:2021ene,Brown:2021rmz,Brown:2021euk,Brown:2022phc}

Circuit complexity is closely related to operator size which measures the number of d.o.f. that an operator can change. In a quantum circuit, operator size is the derivative of the circuit complexity w.r.t. the circuit time~\cite{Susskind:2014jwa}. Inspired by this relation, we adopt the average operator size as the distance function to define complexity. To test whether it is an appropriate definition, a basic criterion is that the complexity geometry should be defined on negatively curved manifolds~\cite{2007quant.ph..1004D,Susskind:2014jwa,Brown:2016wib}. However, we find that the typical sectional curvatures are positive, even if we replace the complexity metric with an arbitrary function of operator size. 

To reconcile this contradiction, we analyze the distribution of sectional curvatures for $N$-qubit system and find that 2-surfaces generated by Hamiltonians of size smaller than the typical size can be negatively curved. Besides the negative curvature, another basic criterion is that the typical sectional curvatures should be at the order of $1/N^2$~\cite{Brown:2016wib}. The complexity metric we choose satisfies this requirement. It can be further shown that our choice is the only metric that satisfies the requirement up to constant corrections.

Regarding the meaning of the analysis on curvature distribution, we should note that some of the directions on the unitary group are not important for a specific physical system. Since a physical Hamiltonian is local or $k$-local, even though the dynamics are quasi-ergodic, many directions induced by some non-local Hamiltonians cannot be detected by the system. In this way, the system is only sensitive to part of the sectional curvatures.

As the distance function of complexity geometry, the operator size measures how many local d.o.f. can be changed by an operator, but it does not measure the number of d.o.f. that are really changed. In general, acting on a state, only part of the operator matters. For example, $\sigma^x\otimes\sigma^z|00\rangle=\sigma^x\otimes I|00\rangle$, where $\sigma^x$ and $\sigma^z$ are Pauli operators, $I$ is the identity. So the operator size should be modified with the knowledge of the state. With this observation, we define state-dependent operator size (SDOS) such that operators with the same action on a given state have the same size. Thus SDOS induces a metric on the Hilbert space with which we can define the complexity geometry of states. Unlike operator complexity geometry, it's not right-invariant and thus more difficult to calculate curvatures. But using Page's theorem~\cite{Page:1993df}, we can show that for $N$-qubit system, 2-surfaces generated by Hamiltonians of size much smaller than $N/2$ acting on typical states are negatively curved.

This paper is organized as follows. In section~\ref{sec:operator size and complexity}, we define operators size and complexity for discrete systems. In section~\ref{sec:positive curvature}, we prove that if the metric on the unitary group is a function of operator size, then typical sectional curvatures are always positive. In section~\ref{sec:sectional curvature}, we analyze the sectional curvature distribution and illustrate that 2-surfaces generated by Hamiltonians of size smaller than the typical size can have negative curvatures. The requirement that sectional curvatures should be at the order of $1/N^2$ constrains the form of the complexity metric up to constant corrections. In section~\ref{sec:state complexity}, state-dependent operator size and state complexity are defined. The curvatures of 2-surfaces generated by Hamiltonians of size much smaller than $N/2$ are negative. Conclusions are in section~\ref{sec:conclusions}. In appendix~\ref{sec:review}, results relevant to sectional curvatures in \cite{2007quant.ph..1004D} are briefly reviewed. In appendix~\ref{app:derivation}, the distribution of sectional curvatures is derived.

\section{Operator size and complexity }
\label{sec:operator size and complexity}
\subsection{Operator size}
Operator size measures how many local d.o.f. an operator can change. It has been analyzed for qubit system~\cite{hosur2016characterizing} and Majorana fermions~\cite{Qi:2018bje}. In this subsection, we define operator size for general discrete systems. 

Given a discrete system, physical operators supported on the subsystem $B$ can be described by a local algebra, $\mathcal{A}_B$, which is the operator algebra~\footnote{An operator algebra is a linear space of operators (including the identity) which is closed under operator multiplication and Hermitian conjugation. Mathematically, it's a von Neumann algebra.} of all the physical operators acting on $B$. By "physical operator", I mean it's gauge-invariant and bosonic~\footnote{Operator algebra can also include fermionic operators. In this case, the algebra is a $Z_2$-graded algebra, which is different from the definition of local algebras in some literature, e.g.~\cite{haag2012local}, where only bosonic operators are included.}. If we can define the measure~\footnote{For example, in the case of qubit model, $|B|$ can be the number of qubits in $B$. Since we are considering discrete systems, we can suppose $|B|\in \mathbb{N}$ without loss of generality.} of $B$, denoted by $|B|$, then we can attribute it to $\mathcal{A}_B$ as a label. Since $\mathcal{A}_B$ is closed under addition and scalar multiplication, $\mathcal{A}_B$ is also a linear space. In the rest of the paper, we denote $\mathcal{A}_B$ as $A_B$ to emphasize this fact. Then we define the linear space $A^k$ as
\begin{equation}
\label{eq:k-local op}
    A^k\equiv \sum_{|B|=k}A_B,
\end{equation}
where $\sum$ means that $A^k$ is spanned by all the vectors in $A_B$'s. We set $A^0\equiv\{I\}$. If $\mathcal{O}\in A^k$, then $\mathcal{O}$ is called a $k$-local operator.

We introduce the inner product between two operators $\mathcal{O}$ and $\mathcal{O}'$
\begin{equation}
\label{eq:op in pro}
    (\mathcal{O}|\mathcal{O}')\equiv |\mathcal{H}|^{-1}\textrm{Tr}(\mathcal{O}^{\dagger}\mathcal{O}'),
\end{equation}
where $|\mathcal{H}|$ is the dimension of the Hilbert space. The symbol $|\mathcal{O})$ indicates that $\mathcal{O}$ is also a vector in the operator space. Then we construct the sequence $\{\Delta A^k\}$ through the orthogonalization
\begin{equation}
\label{eq:orthogonal condition 2}
    A^k=A^{k-1}\oplus \Delta A^k,\quad \Delta A^k \perp A^{k-1} ,
\end{equation}
The projectors $\mathcal{P}_k$
and $\Delta \mathcal{P}_k$ corresponding to $A^k$ and $\Delta A^k$ respectively satisfy
\begin{equation}
  \mathcal{P}_k=\mathcal{P}_{k-1}+\Delta \mathcal{P}_k.
\end{equation}
We define the size operator,
\begin{equation}
\label{eq:size op}
    n\equiv \sum_k k\, \Delta \mathcal{P}_k.
\end{equation}
Then the average size of an operator $\mathcal{O}$ is defined as the expectation value
\begin{equation}
\label{eq:op size}
    \mathcal{S}(\mathcal{O})=\frac{(\mathcal{O}|n|\mathcal{O})}{(\mathcal{O}|\mathcal{O})}.
\end{equation}
A more general form of average is
\begin{equation}
    \mathcal{S}_r(\mathcal{O})=(\frac{(\mathcal{O}|n^r|\mathcal{O})}{(\mathcal{O}|\mathcal{O})})^{\frac{1}{r}}.
\end{equation}
$r$ is a positive number.

Let's take the qubit system for example to illustrate the meaning of the above definitions. Suppose the system is composed of $N$ qubits. If a subsystem $B$ is composed of $k$ qubits, then the number of qubits in $B$ is a natural measure $|B|$ . The local algebra $\mathcal{A}_B$ is composed of all the operators which have trivial actions on the complement of $B$. The linear space $A^k$ can be spanned by all the generalized Pauli operators with a weight smaller than $k+1$,
\begin{equation}
    A^k=\{ \sum_{\{i_a\},\{\alpha_b\}}c_{i_1...i_k}^{\alpha_1...\alpha_k}\sigma_{i_1}^{\alpha_1}...\sigma_{i_k}^{\alpha_k} \,\big| c_{i_1...i_k}^{\alpha_1...\alpha_k} \in \mathbb{C}  ;\, 1\leq i_a \leq N;\, \alpha_b=I, x, y, z\}.
\end{equation}
$\sigma_{i}^{\alpha}$ is the Pauli operator acting on the $i$th qubit. After the orthogonalization, the linear space $\Delta A^k$ is given by
\begin{equation}
    \Delta A^k=\{ \sum_{\{i_a\},\{\alpha_b\}}c_{i_1...i_k}^{\alpha_1...\alpha_k}\sigma_{i_1}^{\alpha_1}...\sigma_{i_k}^{\alpha_k} \,\big| c_{i_1...i_k}^{\alpha_1...\alpha_k} \in \mathbb{C}  ;\, 1\leq i_a \leq N;\, \alpha_b=x, y, z\}.
\end{equation}
The projectors $\mathcal{P}_k$ and $\Delta \mathcal{P}_k$ are given by
\begin{subequations}

\begin{equation}
    \mathcal{P}_k=\sum_{1\leq i_a \leq N} \sum_{\{\alpha_b=I,x, y, z\}}|\sigma_{i_1}^{\alpha_1}...\sigma_{i_k}^{\alpha_k})(\sigma_{i_1}^{\alpha_1}...\sigma_{i_k}^{\alpha_k}|, 
\end{equation}
\begin{equation}
    \Delta \mathcal{P}_k=\sum_{1\leq i_a \leq N} \sum_{\{\alpha_b=x, y, z\}}|\sigma_{i_1}^{\alpha_1}...\sigma_{i_k}^{\alpha_k})(\sigma_{i_1}^{\alpha_1}...\sigma_{i_k}^{\alpha_k}|.
\end{equation}
\end{subequations}
Then the size operator $n$ is
\begin{equation}
    n=\sum_k k\Delta \mathcal{P}_k.
\end{equation}
The size of an operator $\mathcal{O}$ is given by
\begin{equation}
\begin{split}
\mathcal{S}(\mathcal{O})&=\frac{(\mathcal{O}|n|\mathcal{O})}{(\mathcal{O}|\mathcal{O})}
    \\
    &=\frac{1}{(\mathcal{O}|\mathcal{O})}\sum_k \sum_{\{i_a\},\{\alpha_b\}}k\,|(\sigma_{i_1}^{\alpha_1}...\sigma_{i_k}^{\alpha_k}|\mathcal{O})|^2,\quad 1\leq i_a \leq N;\, \alpha_b=x, y, z.
    \end{split}
\end{equation}
For generalized Pauli operators, its size is the number of qubits on which it has nontrivial actions. For an operator $\mathcal{O}$, its size is the average of the size of generalized Pauli operators over the probability distribution $|(\sigma_{i_1}^{\alpha_1}...\sigma_{i_k}^{\alpha_k}|\mathcal{O})|^2$. A more general average is given by
\begin{equation}
    \mathcal{S}_r(\mathcal{O})=[\frac{1}{(\mathcal{O}|\mathcal{O})}\sum_k \sum_{\{i_a\},\{\alpha_b\}}k^r\,|(\sigma_{i_1}^{\alpha_1}...\sigma_{i_k}^{\alpha_k}|\mathcal{O})|^2]^{\frac{1}{r}},\quad 1\leq i_a \leq N;\, \alpha_b=x, y, z.
\end{equation}

\subsection{Operator complexity}
\label{sec:operator complexity}

Size is closely related to complexity. In the circuit model, operator size is the derivative of the circuit complexity w.r.t. the circuit time~\cite{Susskind:2014jwa}. Inspired by this relation, we adopt the operator size defined in the last subsection as the distance function of the operator space, then the complexity is defined as the path length.

\begin{figure}[ht]
\centering  
\includegraphics[scale=0.5]{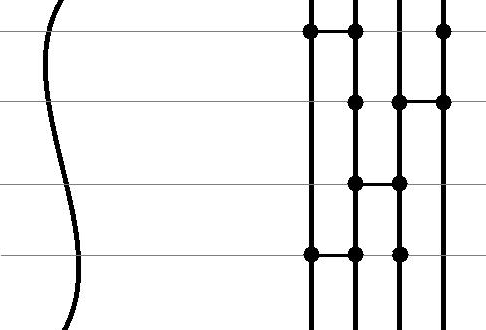}
\caption{\label{fig:op path-circuit} A path in the operator space is on the left, and the corresponding quantum circuit is on the right. Gray lines connect points on the path to the corresponding cross-sections in the circuit. The black dots represent quantum gates.}
\end{figure}

A path in the operator space corresponds to a quantum circuit. Each point on the path is a cross-section of the circuit, see Fig.~\ref{fig:op path-circuit}.
We can regard each cross section as a computation step. At each step, the circuit can change some local d.o.f. which are measured by operator size. So the total number of d.o.f. that a circuit can change during the evolution is the integral (sum) of the operator size along the continuous (discrete) path. If we interpret this total number as the complexity, then the integral measures the complexity of the circuit. In the operator space, the size of an infinitesimal evolution $I-d\mathcal{O}$ is
\begin{equation}
\begin{split}
   \mathcal{S}_r&=[\frac{(I-d\mathcal{O}|n^r|I-d\mathcal{O})}{(I-d\mathcal{O}|I-d\mathcal{O})}]^{\frac{1}{r}}\\
   &=(d\mathcal{O}|n^r|d\mathcal{O})^{\frac{1}{r}}.
\end{split}
\end{equation}
In the second line, we used $n|I)=0$. If we focus on the unitary group, then at the point $U$, the infinitesimal evolution is $I+dUU^{\dagger}$. And the size of it is 
\begin{equation}
    \mathcal{S}_r=(dUU^{\dagger}|n^r|dUU^{\dagger})^{\frac{1}{r}}.
\end{equation}
Since the size is integrated along the path, it should be at the order of $|dUU^{\dagger}|$, which implies $r=2$. Then the complexity of a path $\gamma$ is 
\begin{equation}
\label{eq:op complexity}
    \mathcal{C}=\underset{U\in \gamma}{\int} (dUU^{\dagger}|n^2|dUU^{\dagger})^{\frac{1}{2}}.
\end{equation}

One of the key points to define the complexity is operator distinguishability. We need to distinguish which operators are easy and which operators are hard. This depends on the case we are interested in. In this paper, we use the size to distinguish operators, because we are interested in black holes. The dynamics of black holes is similar to that of $k$-local all-to-all circuits~\cite{Susskind:2014jwa, 
Stanford:2014jda, Roberts:2014isa}. The degrees of freedom in such circuits are de-localized, i.e. each pair of d.o.f. are potentially interacting. For this reason, operators of different localities are not distinguished when defining the complexity. This is different from the usual case like lattice models in which operators of different localities should be distinguished. For example, in a spin chain, two Pauli operators acting on a spin and its neighbor should be distinguished from two Pauli operators acting on two spins far away from each other. Size is not sensitive to the locality and thus is adopted as a candidate to define the complexity.

\section{Positive curvature}
\label{sec:positive curvature}
In the geometric approach, complexity is defined as the distance on the unitary group~\cite{nielsen2005geometric,2007quant.ph..1004D}. In our case, Eq.~\eqref{eq:op complexity} gives the complexity metric
\begin{equation}
\label{eq:op complexity metric}
    d\mathcal{C}^2=(dUU^{\dagger}|n^2|dUU^{\dagger}).
\end{equation} 
Notice that this metric is right-invariant due to $dUV(UV)^{\dagger}=dUU^{\dagger}$ for constant unitary $V$.

To test whether Eq.~\eqref{eq:op complexity} is an appropriate definition, a basic criterion is that curvatures should be negative~\cite{2007quant.ph..1004D,Susskind:2014jwa,Brown:2016wib}. However, from Eq.~\eqref{eq:op complexity metric}, we find the typical sectional curvatures are positive. In fact, it can be further proved that the typical sectional curvatures are always positive, if the metric has the form
\begin{equation}
\label{eq:general metric}
    d\mathcal{C}^2=(dUU^{\dagger}|f(n)|dUU^{\dagger}).
\end{equation}
$f(n)$ is an arbitrary positive-definite function which is called penalty. Eq.~\eqref{eq:general metric} is the usual form of the complexity metric, e.g.~\cite{2007quant.ph..1004D,Brown:2017jil}. To be concrete, we calculate the typical sectional curvatures for the $N$-qubit system.

Randomly choosing two Hamiltonians in the tangent space on the unitary group, $X$ and $Y$, typically they are orthogonal. In general, their Pauli operator expansions contain terms of various sizes. Note that the number of generalized Pauli operators of size $k$ is
\begin{equation}
    D_k=3^k \binom{N}{k}\leq D_{\frac{3}{4}N}.
\end{equation}
So in the large $N$ limit, the expansions are dominated by generalized Pauli operators of size $3N/4$ which implies that approximately $X,\, Y \in \Delta A^{3N/4}$. In Appendix~\ref{app:derivation}, formulas of sectional curvature in various situations are derived. Using Eq.~\eqref{eq:tp sc cur 3}, the typical sectional curvatures are given by
\begin{equation}
    K(\frac{3}{4}N,\frac{3}{4}N)=\frac{1}{2f(\frac{3}{4}N)}.
\end{equation}
We find that typical sectional curvatures are positive and this result doesn't depend on the specific choice of the metric. In other words, the complexity geometry is always defined on almost positively curved manifolds, which contradicts the usual expectation that typical sectional curvatures are negative.

\section{Sectional curvature}
\label{sec:sectional curvature}

To reconcile the contradiction mentioned in the last section, we analyze sectional curvatures distribution and find that the curvatures can become negative for non-typical sections spanned by operators of size smaller than $3N/4$.

\subsection{The distribution of sectional curvatures}

In this subsection, we calculate the curvatures of surfaces generated by $X\in \Delta A^k$ and $Y\in \Delta A^l$. We denote it as $K(k,l)$. Since the typical size of $[X, Y]$ have quite different behaviors when $k,l$ are at different orders of $N$, our calculations are performed in three different situations. For details on the derivation of formulas used in this section, see Appendix~\ref{app:derivation}.

\subsubsection{$k,\,l\sim O(N)$}

\begin{figure}
     \centering
     \begin{subfigure}[b]{0.4\textwidth}
         \centering
         \includegraphics[width=\textwidth]{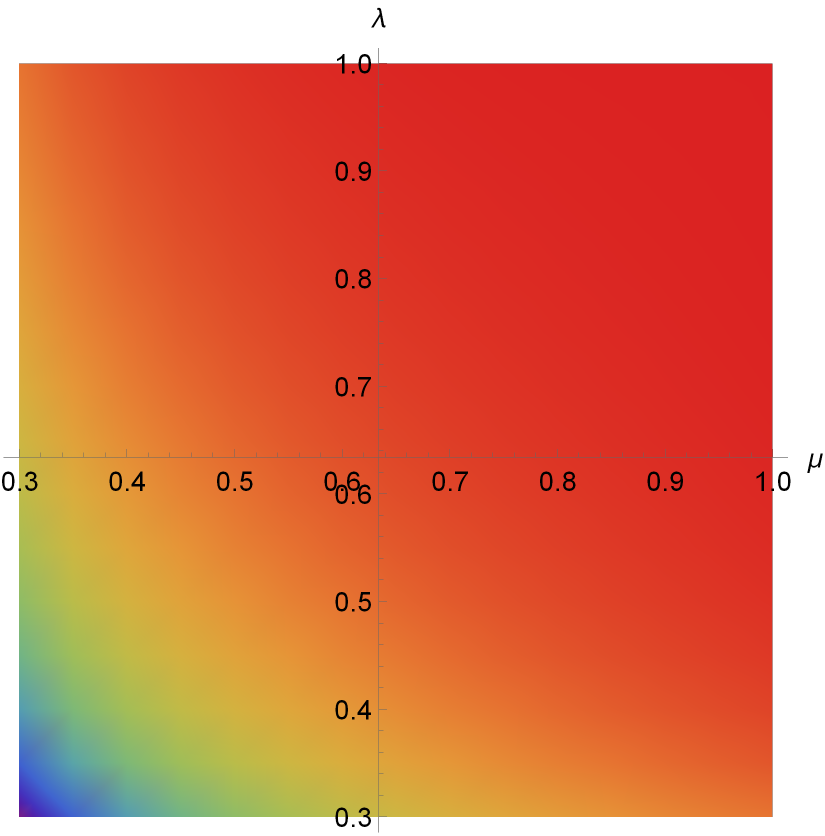}
         \caption{Density plot of $N^2K(k,l)$}
         \label{K1}
     \end{subfigure}
     \hfill
     \begin{subfigure}[b]{0.06\textwidth}
         \centering
         \includegraphics[width=\textwidth]{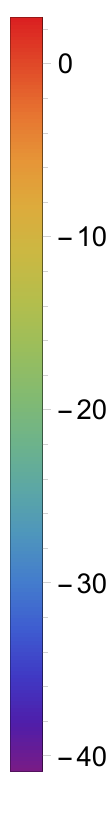}
         
     \end{subfigure}
     \hfill
     \begin{subfigure}[b]{0.5\textwidth}
         \centering
         \includegraphics[width=\textwidth]{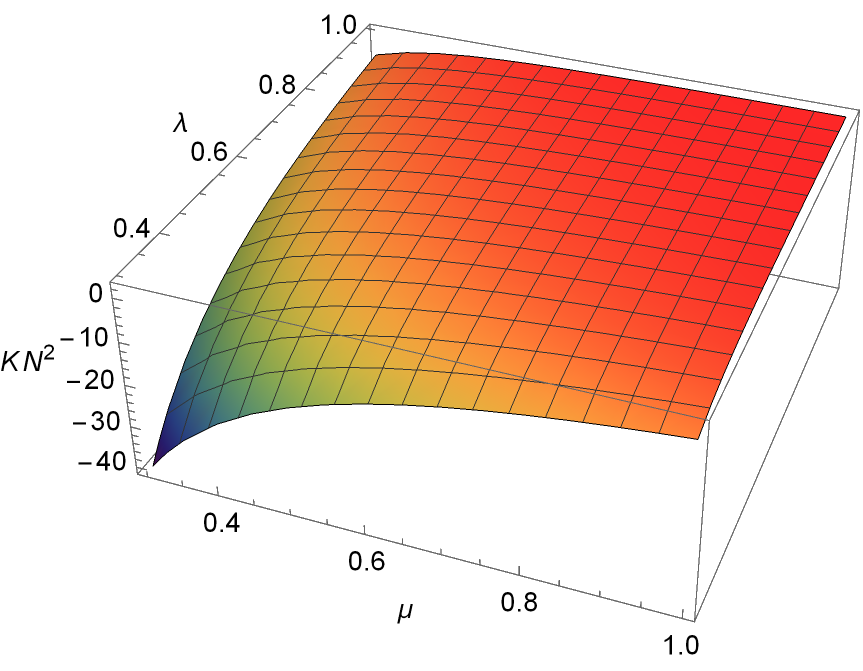}
         \caption{3D plot of $N^2K(k,l)$}
         \label{K-1}
     \end{subfigure}
        \caption{The plot of $N^2K(k,l)$ for $k,\,l\sim O(N)$.}
        \label{fig:typical curvatures}
\end{figure}

\begin{figure}[ht]
\centering  
\includegraphics[scale=0.5]{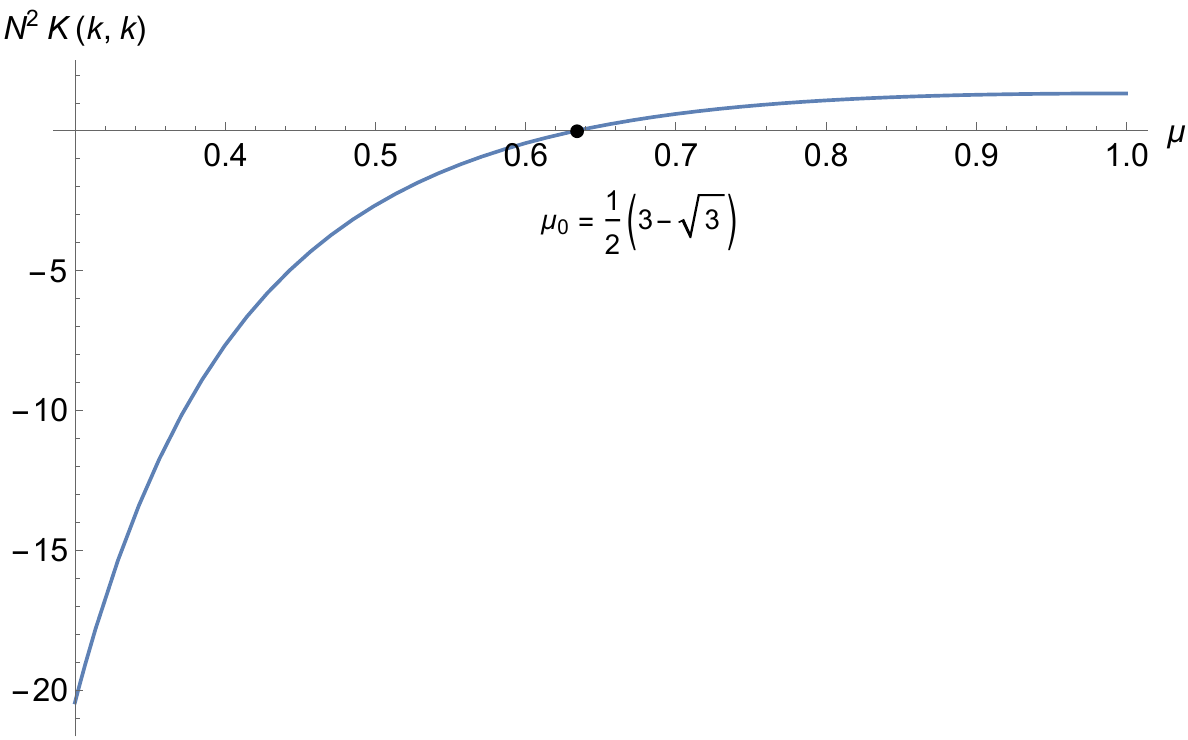}
\caption{\label{fig:curvature-1D} The plot of $N^2K(k,k)$ for $k,\,l\sim O(N)$.}
\end{figure}

Eq.~\eqref{eq:tp sc cur 3} describes the sectional curvatures distribution for general metric Eq.~\eqref{eq:general metric} when $k\sim O(N)$. Applying Eq.~\eqref{eq:tp sc cur 3} to the complexity metric Eq.~\eqref{eq:op complexity metric},
\begin{equation}
\label{eq:cur 1}
\begin{split}
    K(k,l)&=-\frac{1}{2 k^2}-\frac{1}{2 l^2}-\frac{3}{k l}+\frac{4}{N}(\frac{1}{k}+\frac{1}{l})-\frac{8}{3 N^2}
    \\
    &=\frac{1}{N^2}(-\frac{1}{2 \mu^2}-\frac{1}{2 \lambda^2}-\frac{3}{\mu \lambda}+\frac{4}{\mu}+\frac{4}{\lambda}-\frac{8}{3}),
\end{split}
\end{equation}
where $\mu \equiv k/N$ and $\lambda \equiv l/N$. $N^2K(k,l)$ is depicted in Fig.~\ref{fig:typical curvatures} from which we can see
that the curvatures of surfaces generated by operators of smaller size are negative. To be clearer, we set $\mu = \lambda$ in Eq.\eqref{eq:cur 1},
\begin{equation}
    K(k,k)=\frac{4}{N^2}(-\frac{1}{\mu ^2}+\frac{2}{\mu }-\frac{2}{3}).
\end{equation}
We depict it in Fig.~\ref{fig:curvature-1D}. We can observe that curvatures of surfaces generated by operator of size smaller than $\mu_0N$ are negative.

Since $\mu\sim O(1)$ and $\lambda \sim O(1)$, $K(k,l)$ is of order $1/N^2$, which is consistent with the switchback effect of circuit complexity~\cite{Brown:2016wib}. In fact, we will show that to make sure $K(k,l)\sim 1/N^2$, the complexity metric Eq.~\eqref{eq:op complexity metric} is the only appropriate metric up to constant corrections. \footnote{In this paper, we define the operator complexity as the curve length in the operator space. In this case, the complexity of the optimal ``circuit'' preparing an operator is the length of the geodesics connecting the identity and the operator. The following discussion on the metric uniqueness is based on this definition of complexity. However, it can be shown that the geodesics length is proportional to a properly defined on-shell action. So the length formulation of complexity can be replaced by the action formulation~\cite{Brown:2016wib}. In this way, the complexity geometry with typical sectional curvatures $\sim 1/N^2$ in the length formulation can be replaced by the action formulation in which the typical sectional curvatures are proportional to $1/N$ after rescaling the metric appropriately~\cite{Brown:2017jil}.\label{ft:4}}

Consider typical sectional curvatures for the general metric Eq.~\eqref{eq:general metric}, if the asymptotic behavior of the penalty is $f(k)\sim O(k^{\gamma})$, then Eq.~\eqref{eq:tp sc cur 3} gives the asymptotic behavior of curvatures,
\begin{equation}
    K(k,l)\sim O(N^{-\gamma}).
\end{equation}
If $f(k)\sim \textrm{exp}(O(k))$, then we have 
\begin{equation}
    K(k,l)\sim e^{-O(N)}.
\end{equation}
So to be consistent with the switchback effect, $f(k)$ must be at the order of $k^2$. Then in the large $N$ limit, the general form of penalty is
\begin{equation}
    f(k)=ak^2,
\end{equation}
where $a$ is a constant of order one. This penalty is just the penalty in the complexity metric Eq.\eqref{eq:op complexity metric} up to a constant factor.

\subsubsection{$k\sim O(N),\,l\sim O(1)$}

If $k\sim O(N),\,l\sim O(1)$, applying Eq.~\eqref{eq:tp sc cur 3} to the complexity metric Eq.~\eqref{eq:op complexity metric}, we have
\begin{equation}
    K(k,l)=-\frac{1}{2l^2}+O(\frac{1}{N}).
\end{equation}
The curvatures are negative and are in the order of $1$. The leading term only depends on $l$.

\subsubsection{$k,\,l\sim O(1)$}

\begin{figure}[ht]
\centering  
\includegraphics[scale=0.5]{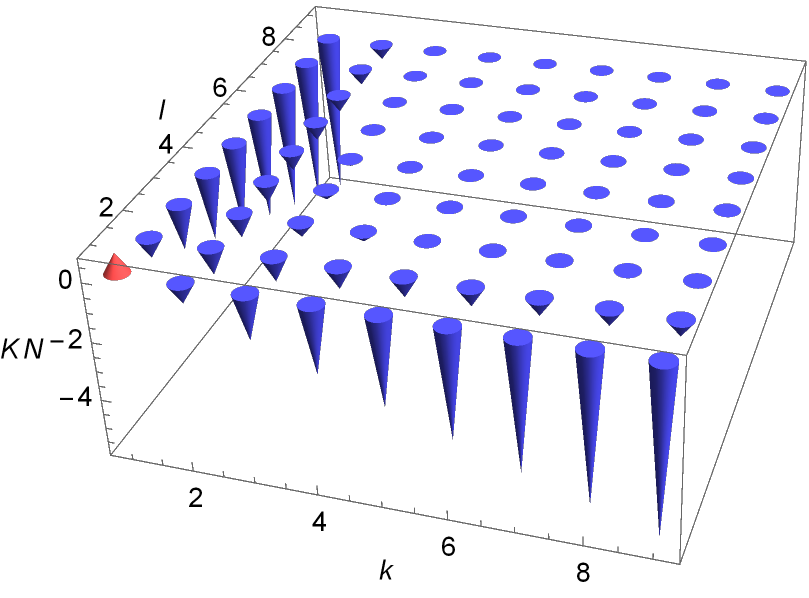}
\caption{\label{fig:curvature-0D} Discrete 3D plot of $NK(k,l)$ for $k,\,l\sim O(1)$. The location of the tip of the cone indicates the value of $NK(k,l)$ at that point. The red cone indicates positive curvature, and blue cones indicate negative curvature. The only red cone is located at (1,\,1) which corresponds to the typical curvature of surfaces generated by 1-local Hamiltonians.}
\end{figure}

Eq.~\eqref{eq:tp cur sc 4} gives the expression of $K(k,l)$ for general metric Eq.~\eqref{eq:general metric} if $k,\,l\sim O(1)$. Applying Eq.~\eqref{eq:tp cur sc 4} to the metric Eq.~\eqref{eq:op complexity metric}, we have,
\begin{equation}
    K(k,l)=\frac{2}{3N}\frac{k!\,l!}{(k+l-2)!} \frac{2 \left(k^2+l^2\right)-3 (k+l-1)^2}{k^2 l^2}.
\end{equation}
The curvatures are at the order of $1/N$.~\footnote{As mentioned in footnote~\ref{ft:4}, if sectional curvatures are of order $1/N$ and negative, the switchback effect can still be reproduced in the action formulation of complexity geometry~\cite{Brown:2017jil}.} Set $l=1$, we have
\begin{equation}
    K(k,1)=\frac{2k}{3N} \frac{2-k^2}{k^2}.
\end{equation}
If $k\geq 2$, then $K(k,1)< 0$. If $(k,l)\in [2,\infty)\times[2,\infty)$, direct calculation shows that the difference is always positive,
\begin{subequations}
\begin{equation}
K(k+1,l)-K(k,l)> 0,
\end{equation}

\begin{equation}
K(k,l+1)-K(k,l)> 0.
\end{equation}
\end{subequations}
Since for large $k,l$ the curvatures are close to zero,
\begin{equation}
    \underset{k,l\rightarrow \infty}{\textrm{lim}}K(k,l)=0,
\end{equation}
we conclude that for $k,\,l\sim O(1)$, the curvature $K(k,l)$ is always negative except for the case of $k=l=1$. See Fig.~\ref{fig:curvature-0D}.

\subsection{Other penalties}

Besides Eq.~\eqref{eq:op complexity metric}, other penalties were also proposed~\cite{2007quant.ph..1004D,Brown:2017jil}. In the following, based on results derived in Appendix~\ref{app:derivation}, we discuss some important features of these geometries.

\subsubsection{Dowling-Nielsen penalty}

The authors of \cite{2007quant.ph..1004D} choose the penalty in Eq.~\eqref{eq:general metric} as
\begin{equation}
    f(k)=
    \left\{
\begin{array}{c}
     1, \quad k\leq 2, \\
    q, \quad 2<k. \\
\end{array}
\right.
\end{equation}
$q$ is a positive constant. If $k,l>2$, Using Eq.~\eqref{eq:tp sc cur 2}, we get
\begin{equation}
    K(k,l)=\frac{D_{k,l}}{D_k D_l\, q},
\end{equation}
where $D_k$ and $D_l$ are dimensions of $\Delta A^k$ and $\Delta A^l$ respectively. $D_{k,l}$ is the number of anticommuting pairs $(\rho,\sigma)\in \Delta A^k \times \Delta A^l $. If $k\leq2$ and $l>2$, then Eq.~\eqref{eq:tp sc cur 3}\eqref{eq:tp cur sc 4} imply that
\begin{equation}
    K(k,l)=
    \left\{
\begin{array}{c}
     \frac{2}{3N}\frac{k!\,l!}{(k+l-2)!}(\frac{2}{q}-1), \quad l\sim O(1), \\
    \frac{1}{q}-\frac{1}{2}, \quad l\sim O(N). \\
\end{array}
\right.
\end{equation}
It's negative for $q>2$. So sectional curvatures are almost always positive except for surfaces generated by 1-local or 2-local Hamiltonians, which contradicts the negative curvature criterion. Similar issues were discussed in \cite{Auzzi:2020idm}.

\subsubsection{Brown-Susskind penalty}

The authors of \cite{Brown:2017jil} choose the penalty in Eq.~\eqref{eq:general metric} as
\begin{equation}
    f(k)=
    \left\{
\begin{array}{c}
     1, \quad k\leq 2, \\
    c\,4^{k-2}, \quad 2<k. \\
\end{array}
\right.
\end{equation}
$c$ is a constant of order one. If $k,l \sim O(N)$ and $k,l<3N/4$, Eq.~\eqref{eq:tp sc cur 3} implies that
\begin{equation}
\label{eq:susskind curvature}
    K(k,l)\approx-\frac{3}{2c}4^{-\frac{4kl}{3N}}\sim e^{-O(N)},
\end{equation}
So the curvature length is exponentially large. According to the argument in \cite{Brown:2016wib}, the curvature length controls the scrambling time, then Eq.~\eqref{eq:susskind curvature} implies exponentially large scrambling time, which contradicts with the fact that random circuit is a fast scrambler.

\section{State complexity}
\label{sec:state complexity}

\subsection{State-dependent operator size}
Operator size defined in Eq.~\eqref{eq:op size} measures how many local d.o.f. an operator can change, but it doesn't measure the number of d.o.f. that are really changed for the operator acting on a given state.  For example, consider the operator $\mathcal{O}=\sigma^x\otimes\sigma^z$ and the state $|00\rangle$. The size of $\mathcal{O}$ is 2. On the other hand, since $\mathcal{O}|00\rangle=|10\rangle$, only one d.o.f. is changed. 

With the knowledge of the state, the operator size should be modified. In the following, we denote the given state as $|\psi\rangle$. We first construct the subspace
\begin{equation}
    \mathcal{H}_k\equiv \{\mathcal{O}|\psi\rangle\,\big| \, \mathcal{O}\in A^k \},
\end{equation}
where $A^k$ is the linear space defined in Eq.~\eqref{eq:k-local op}. $\mathcal{H}_k$ is spanned by all the states that can be generated by operators acting on $|\psi\rangle$ with the restriction that the measure of support of these operators is equal to or smaller than $k$. For a state $|\phi\rangle$ that satisfy $|\phi\rangle\in \mathcal{H}_{k} $ and $|\phi\rangle \notin \mathcal{H}_{k-1}$, note that the overlap between $|\phi\rangle$ and states in $\mathcal{H}_{k-1}$ is not zero generally, which means that $|\phi\rangle$ can be decomposed as
\begin{equation}
    |\phi\rangle=P_{k-1}|\phi\rangle+(I-P_{k-1})|\phi\rangle,
\end{equation}
where $P_{k-1}$ is the projector corresponding to $\mathcal{H}_{k-1}$. $P_{k-1}|\phi\rangle$ is the redundant component of $|\phi\rangle$ which can be generated without $k$-local operators acting on $|\psi\rangle$. In general, $\mathcal{H}_{k}$ can be decomposed as 
\begin{equation}
\label{eq:orthogonal condition}
\Delta \mathcal{H}_k\oplus\mathcal{H}_{k-1}\equiv\mathcal{H}_k,
\quad
\Delta \mathcal{H}_k \perp\mathcal{H}_{k-1},
\quad 1\leq k\leq N_{\psi }.
\end{equation}
$N_{\psi}$ is the minimal value of $k$ such that $\mathcal{H}_{k+1}=\mathcal{H}_{k}$. We set $\mathcal{H}_{-1}\equiv \emptyset$. The meaning of $\Delta\mathcal{H}_k$ is that we only need $k$-local operators acting on $|\psi\rangle$ to generate $\Delta \mathcal{H}_k$ which cannot be generated by $(k-1)$-local operators, and each state in $\Delta \mathcal{H}_k$ can be perfectly distinguished from states that can be generated by $(k-1)$-local operators acting on $|\psi\rangle$. Then we can say that there are $k$ local d.o.f. that are changed when $|\psi\rangle$ evolves to a state in $\Delta\mathcal{H}_k$. For this reason, we call $\Delta\mathcal{H}_k$ the $k$-local subspace. We have the orthogonalization decomposition
\begin{equation}
\label{space decomposition}
\mathcal{H}=\bigoplus_{k=0}^{N_{\psi}} \Delta \mathcal{H}_k,
\end{equation}
where $\mathcal{H}$ is the finite dimensional Hilbert space of the whole system. For a general operator $\mathcal{O}$, we have
\begin{equation}
\mathcal{O}|\psi\rangle=\sum _{k=0}^{N_{\psi }} \Delta P_k\mathcal{O}| \psi\rangle,
\end{equation}
where $\Delta P_k$ is the projector corresponding to $\Delta\mathcal{H}_k$ and satisfies $\Delta P_k=P_k-P_{k-1}$. Then we define the state-dependent operator size (SDOS) as the average number of local d.o.f. that are changed by $\mathcal{O}$ acting on $|\psi\rangle$,
\begin{equation}
\mathcal{S}_1(\mathcal{O},|\psi\rangle)=\sum _{k=1}^{N_{\psi }} k \left\langle \psi \left|\mathcal{O}^{\dagger}\Delta P_k\mathcal{O}\right|\psi \right\rangle,
\end{equation}
and we can consider a more general average
\begin{equation}
\mathcal{S}_r(\mathcal{O},|\psi\rangle)=(\sum _{k=1}^{N_{\psi }} k^r \left\langle \psi \left|\mathcal{O}^{\dagger}\Delta P_k\mathcal{O}\right|\psi \right\rangle)^{\frac{1}{r}},
\end{equation}
where the normalization $\langle\psi|\mathcal{O}^{\dagger}\mathcal{O}|\psi\rangle=1$ is assumed. Note that if $\mathcal{O}|\psi\rangle=\mathcal{O}'|\psi\rangle$, we have
\begin{equation}
    \mathcal{S}_r(\mathcal{O},|\psi\rangle)=\mathcal{S}_r(\mathcal{O}',|\psi\rangle).
\end{equation}
So SDOS can be regarded as a function of initial state $|\psi\rangle$ and finial state $|\phi\rangle$,
\begin{equation}
\mathcal{S}_r(|\phi\rangle,|\psi\rangle)=(\sum _{k=1}^{N_{\psi }} k^r \left\langle \phi \left|\Delta P_k\right|\phi \right\rangle)^{\frac{1}{r}}.
\end{equation}
It is convenient to define the state-dependent size operator $n_{\psi}$,
\begin{equation}
\label{eq:SDOS operator}
n_{\psi}\equiv \sum _{k=1}^{N_{\psi }} k\,\Delta P_k ,
\end{equation}
then we have
\begin{equation}
\label{eq:r SDOS}
\mathcal{S}_r(|\phi\rangle,|\psi\rangle)=\langle \phi \left|(n_{\psi})^r|\phi \right\rangle ^{\frac{1}{r}}.
\end{equation}

\subsection{State complexity}

The evolution of the system corresponds to a path in the Hilbert space, each infinitesimal segment of the path can be regarded as a computation step. For the same reason mentioned in Sec.~\ref{sec:operator complexity}, if we define the complexity as the total number of changed d.o.f. during the evolution, the complexity of the path $\gamma$ can be expressed as the integral
\begin{equation}
\label{eq:state complexity}
    \mathcal{C}=\underset{|\psi\rangle\in \gamma}{\int}  \mathcal{S}_2(|\psi\rangle+|d\psi\rangle,|\psi\rangle).
\end{equation}
We have set $r=2$ to make sure that $\mathcal{S}_r(|\psi\rangle+d|\psi\rangle,|\psi\rangle)\sim O(||d\psi\rangle|)$. If we regard the state complexity $\mathcal{C}$ as the path length, then Eq.~\eqref{eq:state complexity} induces a metric on the Hilbert space,
\begin{equation}
\label{eq:state metric}
    d\mathcal{C}^2=\langle d\psi|(n_{\psi})^2|d\psi\rangle.
\end{equation}
To exhibit chaotic behaviors with large degrees of freedom, the Hilbert space should be negatively curved,  but it's hard to compute $n_{\psi}$ generally~\footnote{Note that if we flatten the spectrum of $n_{\psi}$, i.e.
\begin{equation*}
    \underset{m\rightarrow 0}{\lim}\langle d\psi|(n_{\psi})^m|d\psi\rangle=\langle d\psi|d\psi\rangle-\langle d\psi|\psi\rangle\langle\psi|d\psi\rangle,
\end{equation*}
the complexity metric reduces to the Fubini-Study metric. So the complexity metric here is a deformation of the Fubini-Study metric such that displacements induced by Hamiltonians of different sizes can be distinguished.}. However, it can be shown that for typical states, the curvatures of surfaces generated by local Hamiltonians are negative.

According to Page's theorem~\cite{Page:1993df}, if a system is divided into two parts, $B_1$ and $B_2$, the entanglement entropy of a typical state is given by \begin{equation}
    S=\log (|B_1|)-\frac{1}{2}\frac{|B_1|}{|B_2|}+O(\frac{1}{|B_1||B_2|}).
\end{equation}
$|B_1|$ and $|B_2|$ are Hilbert space dimensions of $B_1$ and $B_2$ respectively. $|B_2|$ is large and $|B_1|<|B_2| $. If $|B_1|\ll|B_2|$, then a typical state is very close to the maximally mixed state after tracing over $B_2$. For two product operators $o_k$ and $o_l$ of size $k$ and $l$ respectively, if $k+l\ll N/2$, then the expectation value w.r.t. a typical state, $|\psi\rangle$, is 
\begin{equation}
    \langle\psi|o_k^{\dagger}o_l|\psi\rangle \approx |\mathcal{H}|^{-1} \textrm{Tr}(o_k^{\dagger}o_l)=(o_k|o_l).
\end{equation}
Since a $k$-local operator can be expressed as the sum of operators of size equal to or smaller than $k$, we have
\begin{equation}
\label{eq:Page}
    \langle\psi|\mathcal{O}_k^{\dagger}\mathcal{O}_l|\psi\rangle \approx |\mathcal{H}|^{-1} \textrm{Tr}(\mathcal{O}_k^{\dagger}\mathcal{O}_l)=(\mathcal{O}_k|\mathcal{O}_l),
\end{equation}
where $|\mathcal{H}|$ is the Hilbert space dimension, $\mathcal{O}_k \in A^k$, and $\mathcal{O}_l \in A^l$.

Starting from a typical state $|\psi\rangle$, we calculate the length of an infinitesimal displacement induced by a $k$-local Hamiltonian, $H$, with $k\ll N/2$.
\begin{equation}
\label{eq:54}
    d\mathcal{C}=\langle \psi|H(n_{\psi})^2H|\psi\rangle^{\frac{1}{2}}dt.
\end{equation}
$t$ is the circuit time parameter. Using Eq.~\eqref{eq:SDOS operator}, we have
\begin{equation}
\label{eq:55}
    \langle \psi|H(n_{\psi})^2H|\psi\rangle=\sum _{l=1}^{k} l^2\,\langle \psi|H\Delta P_l H|\psi\rangle.
\end{equation}
Since $\Delta P_l=P_l-P_{l-1}$, we just need to calculate $\langle \psi|H P_l H|\psi\rangle$. $P_l H|\psi\rangle$ is the projection of $H|\psi\rangle$ onto $\mathcal{H}_l$, so there exists an operator $\mathcal{O}_l\in A^l$ such that
 \begin{equation}
     P_l H|\psi\rangle=\mathcal{O}_l|\psi\rangle.
 \end{equation}
Then
 \begin{equation}
 \label{eq:57}
    \langle \psi|H P_l H|\psi\rangle=\langle \psi|H\mathcal{O}_l|\psi\rangle \approx (H|\mathcal{O}_l).
 \end{equation}
In the second equality, we used Eq.~\eqref{eq:Page}. As the projection, $\mathcal{O}_l|\psi\rangle$ minimizes the distance~\footnote{Here the distance refers to the distance between two vectors in a linear space. It's defined through the inner product of quantum states. This should not be confused with the geodesics length if we regard the Hilbert space as a manifold with the metric defined by Eq.~\eqref{eq:state metric}.} from states in $\mathcal{H}_l$ to $H|\psi\rangle$, $||(\mathcal{O}_l-H)|\psi\rangle||$. Note that
\begin{equation}
\begin{split}
    ||(\mathcal{O}_l-H)|\psi\rangle||
    &=\langle\psi|(\mathcal{O}_l^{\dagger}-H)(\mathcal{O}_l-H)|\psi\rangle^{\frac{1}{2}}
    \\
    & \approx
    (\mathcal{O}_l-H|\mathcal{O}_l-H)^{\frac{1}{2}}.
\end{split}
\end{equation}
This is also the distance between $\mathcal{O}_l$ and $H$ with the inner product defined by Eq.~\eqref{eq:op in pro}. It implies that $O_l$ minimizes the distance from operators in $A^l$ to $H$, and thus $\mathcal{O}_l$ is the projection of $H$ onto $A^l$,
\begin{equation}
\label{eq:59}
    \mathcal{P}_l| H)=|\mathcal{O}_l).
\end{equation}
 So the line element in Eq.~\eqref{eq:54} can be expressed as
 \begin{equation}
 \label{eq:60}
 \begin{split}
     d\mathcal{C}^{2}&=\langle \psi|H(n_{\psi})^2H|\psi\rangle dt^2
     \\
     &=
     \sum _{l=1}^{k} l^2\,\langle \psi|H\Delta P_l H|\psi\rangle dt^2
     \\
     &\approx
     \sum _{l=1}^{k} l^2\,[(H|\mathcal{O}_l)-(H|\mathcal{O}_{l-1})] dt^2
     \\
     &=
     \sum _{l=1}^{k} l^2\,(H|\Delta\mathcal{P}_l |H) dt^2
     \\
     &=
     (H|n^2|H)dt^2
     \\
     &=
     (dUU^{\dagger}|n^2|dUU^{\dagger}).
 \end{split}
 \end{equation}
 In the second equality, we used Eq.~\eqref{eq:55}. In the third equality, we used Eq.~\eqref{eq:57}. In the forth euality, we used Eq.~\eqref{eq:59}. In the fifth equality, we used Eq.~\eqref{eq:size op}. In the sixth euality, we used Eq.~\eqref{eq:op complexity metric} and suppose that $H$ is the generator of the unitary circuit $U$, i.e.
 \begin{equation}
     dUU^{\dagger}=-iHdt.
 \end{equation}
Eq.~\eqref{eq:60} implies that, on surfaces generated by local Hamiltonians, the geometry described by Eq.~\eqref{eq:state metric} is approximately equivalent to the geometry described by Eq.~\eqref{eq:op complexity metric}. So the results in Sec.~\ref{sec:sectional curvature} imply that, in the Hilbert space, surfaces generated by Hamiltonians of size much smaller than $N/2$ acting on typical states~\footnote{To apply the Page's theorem, we need to assume the state is typical. Some ordered states are also of importance like the ground states of a local Hamiltonian, but the analysis on curvatures here doesn't apply to such cases.} are negatively curved.

\section{Conclusions}
\label{sec:conclusions}
In this work, we generalize operator size to arbitrary discrete systems and choose it as the distance function on operator space to define the complexity of the circuit generating an operator. We prove that typical sectional curvatures of the unitary group with the operator size as the distance function are always positive, even if we replace the distance function with an arbitrary function of size. This seems inconsistent with chaotic behaviors of circuit complexity~\footnote{In \cite{Brown:2016wib}, it was shown that there is an analogy between circuit complexity and the geodesics length in a negatively curved space. However, it is not known how the sign of the typical sectional curvatures is related to the chaotic behaviors in general cases.}. To reconcile this contradiction, we analyze the sectional curvatures distribution for the $N$-qubit system and find that 2-surfaces generated by Hamiltonians of size smaller than typical size are still negatively curved. Based on the curvature analysis, we show that to make sure the typical sectional curvatures are of order $1/N^2$, the complexity metric defined in Sec.~\ref{sec:positive curvature} is the only choice in the length formulation of complexity geometry up to constant corrections. To further define the state complexity, we note that with the knowledge of states, the operator size should be modified due to the redundant action of operators. With this consideration, we define the state-dependent operator size (SDOS) and use it as the distance function on the Hilbert space to define the state complexity. Using Page's theorem, we show that 2-surfaces generated by Hamiltonians of size much smaller than $N/2$ acting on typical states are also negatively curved.

Except as the distance function of complexity geometry, another motivation to define the SDOS is that operator size itself is also an important quantity in the study of quantum chaos~\cite{Roberts:2018mnp,Qi:2018bje,hosur2016characterizing} and quantum gravity~\cite{Susskind:2018tei,Susskind:2019ddc,Brown:2018kvn,Lensky:2020ubw}. Especially, it was conjectured that operator size has a gravitational dual which is related to radial momentum~\cite{Susskind:2019ddc}. However, there is an ambiguity that operators of different sizes can correspond to the same particle in the bulk due to the quantum error correction property of AdS/CFT~\cite{Almheiri:2014lwa}, so which operator size is dual to the momentum? To answer the question, we should notice that these operators have the same action on the state. The operator size measures the number of local d.o.f. that the operator can change but doesn't measure how many local d.o.f. are really changed. The latter is measured by SDOS. So although different operator sizes are related to the same radial momentum, these operators have the same state-dependent operator size. Then the ambiguity can be fixed. So it will be interesting to further study the relation between the SDOS and momentum.

After this work is finished, I found that the author of \cite{Mousatov:2019xmc} also proposed a form of state-dependent operator size with the same spirit that operators should have the same size if they have the same action on the state. The author defined it as the minimal size the operator can have among a class of operators with action on the state. The gravity dual is also discussed for the SYK model and holographic 2D CFT.

\section*{Acknowledgments}
I am would like to thank Ling-Yan Hung and Xiao-Liang Qi for helpful discussions and valuable insights. I am also grateful to Yong-Shi Wu for his long-term help in this tough period. I acknowledge the support of the Shenzhen Institute for Quantum Science and Engineering,
Southern University of Science and Technology and Fudan University.

\bibliographystyle{unsrt}  
\bibliography{references}

\appendix

\section{A review of penalty geometry}
\label{sec:review}

This subsection briefly reviews the results in \cite{2007quant.ph..1004D} relevant to sectional curvature. The authors defined a general right-invariant metric on SU$(2^N)$. The metric components in the Hamiltonian representation are given by
\begin{equation}
\label{eq:penalty metric}
    (H|J)_{\mathcal{C}}\equiv (H|\mathcal{G}|J)=2^{-N}\textrm{Tr}(H\mathcal{G}(J)).
\end{equation}
$H$ and $J$ are Hamiltonians specifying two directions. $\mathcal{G}$ is a superoperator that is positive-definite. The Riemann curvature tensor is defined by
\begin{equation}
    R(W,X,Y,Z)\equiv (\nabla_W\nabla_XY-\nabla_X\nabla_W Y-\nabla_{i[W,X]}Y|Z)_{\mathcal{C}},
\end{equation}
where $W,X,Y,Z$ are operators. The sectional curvature in the tangent plane spanned
by two orthogonal operators $X$ and $Y$ is
defined by
\begin{equation}
    K(X,Y)\equiv \frac{ R(X,Y,Y,X)}{(X|X)_{\mathcal{C}}(Y|Y)_{\mathcal{C}}}.
\end{equation}
Then direct calculation shows that
\begin{equation}
\label{eq:general sectional curvature}
    K(X,Y)=-\frac{3}{4}|i[X,Y]|^2_{\mathcal{C}}+|\mathcal{B}_S(X,Y)|^2_{\mathcal{C}}+(i[X,Y]|\mathcal{B}_A(X,Y))_{\mathcal{C}},
\end{equation}
where the norm is defined by $|\bullet|^2_{\mathcal{C}}\equiv (\bullet|\bullet)_{\mathcal{C}}$. $\mathcal{B}_S$ and $\mathcal{B}_A$ are defined by
\begin{equation}
    \mathcal{B}_S(X,Y)\equiv \frac{1}{2}(\mathcal{B}(X,Y)+\mathcal{B}(Y,X)),
\end{equation}
\begin{equation}
    \mathcal{B}_A(X,Y)\equiv \frac{1}{2}(\mathcal{B}(X,Y)-\mathcal{B}(Y,X)).
\end{equation}
$\mathcal{B}(X,Y)$ is defined as $\mathcal{G}^{-1}(i[\mathcal{G}(X),Y])$.

If we set $\mathcal{G}$ as $n^2$ in Eq.~\eqref{eq:penalty metric}, then we recover the operator complexity metric Eq.~\eqref{eq:op complexity metric}.

\section{Derivation of the curvatures distribution}
\label{app:derivation}

\subsection{Main formulas}
Suppose the complexity metric is a function of operator size and takes the following form
\begin{equation}
    (X|Y)_{\mathcal{C}}=f(k) \delta_{XY},
\end{equation}
where $X\in \Delta A^k$, $Y \in \Delta A^l$ and $f(k)$ is a positive functio of $k$ called penalty. Then $\mathcal{B}(X,Y)$ is given by
\begin{equation}
    \mathcal{B}(X,Y)=f(k)\mathcal{G}^{-1}(i[X,Y]).
\end{equation}
$X$ and $Y$ can be expanded in the basis of generalized Pauli operators
\begin{subequations}
\label{eq:pauli expansion}
\begin{equation}
X=\sum_{\{i_a\},\{\alpha_b\}}X_{i_1...i_k}^{\alpha_1...\alpha_k}\sigma_{i_1}^{\alpha_1}...\sigma_{i_k}^{\alpha_k},\quad 1\leq i_a \leq N,\, \alpha=x,y,z.
    \end{equation}
    
    \begin{equation}
Y=\sum_{\{i_a\},\{\alpha_b\}}Y_{i_1...i_l}^{\alpha_1...\alpha_l}\sigma_{i_1}^{\alpha_1}...\sigma_{i_l}^{\alpha_l},\quad 1\leq i_a \leq N,\, \alpha=x,y,z.
    \end{equation}
\end{subequations}
In general, $|[X,Y])$ is not an eigenvector of $\mathcal{G}$, but in the large $N$ limit, most of terms in its Pauli operator expansion have the same size, denoted as $\mathcal{S}(k,l)$. So the typical value of $\mathcal{B}_(X,Y)$ is
\begin{equation}
\label{eq:B(X,Y)}
    \mathcal{B}(X,Y)=\frac{f(k)}{f(\mathcal{S}(k,l))}i[X,Y].
\end{equation}
For the same reason, using Eq.~\eqref{eq:penalty metric}, the typical value of $|[X,Y]|^2_{\mathcal{C}}$ is
\begin{equation}
\label{eq:[X,Y]}
   |[X,Y]|^2_{\mathcal{C}}= 2^{-N}f(\mathcal{S}(k,l))\textrm{Tr}([X,Y][Y,X]).
\end{equation}
Inserting Eq.~\eqref{eq:B(X,Y)} \eqref{eq:[X,Y]} into Eq.~\eqref{eq:general sectional curvature}, the typical sectional curvature is given by
\begin{equation}
\label{eq:tp sc cur 1}
    K(X,Y)=2^N\frac{\textrm{Tr}(\frac{1}{4}[X,Y][Y,X])}{\textrm{Tr}(X^2)\textrm{Tr}(Y^2)}\frac{-3f(\mathcal{S}(k,l))+2(f(k)+f(l))}{f(k)f(l)}.
\end{equation}
Then we need to claculate the typical value of the commutator square. For convenience, we rewrite Eq.~\eqref{eq:pauli expansion} as 
\begin{subequations}

\begin{equation}
X=\sum_{I}X_I \Gamma_I,
    \end{equation}
    
\begin{equation}
Y=\sum_{I}Y_I \Gamma_I,
    \end{equation}
\end{subequations}
where $\Gamma$ stands for the generalized operator and $I$ stands for the indexes $\{i_a\}$ and $\{\alpha_b\}$. 
Then the commutator square can be written as
\begin{equation}
    2^N\frac{\textrm{Tr}(\frac{1}{4}[X,Y][Y,X])}{\textrm{Tr}(X^2)\textrm{Tr}(Y^2)}=2^N\frac{\frac{1}{4}\sum_{I,J}|X_I|^2|Y_J|^2\textrm{Tr}([\Gamma_I,\Gamma_J][\Gamma_J,\Gamma_I])}{(\sum_I |X_I|^2\textrm{Tr}(\Gamma_I^2))(\sum_J |Y_J|^2\textrm{Tr}(\Gamma_J^2))}.
\end{equation}
Since the generalized Pauli operator satisfy the relation $\Gamma_I^2=I$ and
\begin{equation}
\textrm{Tr}([\Gamma_I,\Gamma_J][\Gamma_J,\Gamma_I])= \left\{
\begin{array}{c}
2^{N+2},\quad [\Gamma_I,\Gamma_J]\neq 0,\\
0, \quad [\Gamma_I,\Gamma_J]= 0.
\end{array}
\right.
\end{equation}
the commutator square is given by
\begin{equation}
    2^N\frac{\textrm{Tr}(\frac{1}{4}[X,Y][Y,X])}{\textrm{Tr}(X^2)\textrm{Tr}(Y^2)}=\frac{\sum_{[\Gamma_I,\Gamma_J]\neq 0}|X_I|^2|Y_J|^2}{(\sum_I |X_I|^2)(\sum_J |Y_J|^2)}.
\end{equation}
Suppose $X_I$ and $Y_J$ are random variables in some ensemble and the second moment is $c$, then the typical value of the commutator square is given by
\begin{equation}
\label{eq:commutator square}
\begin{split}
    2^N\frac{\textrm{Tr}(\frac{1}{4}[X,Y][Y,X])}{\textrm{Tr}(X^2)\textrm{Tr}(Y^2)}&=\frac{\sum_{[\Gamma_I,\Gamma_J]\neq 0}c^2}{(\sum_I c)(\sum_J c)}\\
    &=\frac{D_{k,l}}{D_k D_l}.
\end{split}
\end{equation}
$D_k$ and $D_l$ are dimensions of $\Delta A^k$ and $\Delta A^l$ respectively. $D_{k,l}$ is the number of anticommuting pairs $(\rho,\sigma)\in \Delta A^k \times \Delta A^l $. Insert Eq.~\eqref{eq:commutator square} into Eq.~\eqref{eq:tp sc cur 1},
\begin{equation}
\label{eq:tp sc cur 2}
   K(k,l)=\frac{D_{k,l}}{D_k D_l f(k)f(l)}(-3f(\mathcal{S}(k,l))+2(f(k)+f(l))).
\end{equation}
Here $K(k,l)$ stands for the typical value of $K(X,Y)$.

\subsection{Commutator size}
To apply Eq.~\eqref{eq:tp sc cur 2}, we need to know  $D_{k,l}$ and $\mathcal{S}(k,l)$ in the large $N$ limit. W calculate them for two random Hamiltonians in two different situations. In the following, $\rho \in \Delta A ^k$ and $\sigma \in \Delta A^l$ represent two random generalized Pauli operators.

\subsubsection{$k\sim O(N)$ }

Given a fixed $\rho$, randomly picking up one qubit from the $N$-qubit system, the probability that $\rho$ has nontrivial action on this qubit is $k/N$. So the typical size of the overlap between the domain of $\rho$ and $\sigma$ is
\begin{equation}
    |\mathcal{D}_{\rho}\cap \mathcal{D}_{\sigma}|=\frac{kl}{N}.
\end{equation}
$\mathcal{D}_{\rho}$ is the set of qubits on which $\rho$ has nontrivial action. $|\mathcal{D}_{\rho}|$ is the number of qubits in $\mathcal{D}_{\rho}$. Randomly choosing two Pauli operators acting on one qubit, there are nine combinations. Only two-thirds of them are not the identity. So the typical size of $\rho \sigma$ is 
\begin{equation}
\begin{split}
    |\rho \sigma|_{\mathcal{C}}&=\frac{2}{3}|\mathcal{D}_{\rho}\cap \mathcal{D}_{\sigma}|+(|\mathcal{D}_{\rho}|-|\mathcal{D}_{\rho}\cap \mathcal{D}_{\sigma}|)+(|\mathcal{D}_{\sigma}|-|\mathcal{D}_{\rho}\cap \mathcal{D}_{\sigma}|)\\
    &=k+l-\frac{4kl}{3N}.
\end{split}
\end{equation}
Since $|\rho \sigma|_{\mathcal{C}}=|\sigma \rho|_{\mathcal{C}}$,  if $[\rho,\sigma]\neq 0$,
the typical size of $[\rho, \sigma]$ is $|\rho \sigma|_{\mathcal{C}}$. If they commute, then the size is 0. Two Pauli operators acting on one qubit always commute or anticommute, so typically, $[\rho,\sigma]\neq 0$ implies that $\frac{2}{3}|\mathcal{D}_{\rho}\cap \mathcal{D}_{\sigma}|$ is an odd number and vice versa. Then the typical size of $[\rho,\sigma]$ is given by
\begin{equation}
\label{eq:Pauli commutator size 1}
    \mathcal{S}(\rho,\sigma)=
    \left\{
\begin{array}{c}
     k+l-\frac{4kl}{3N}, \quad \textrm{for almost one half of pairs }(\rho,\sigma), \\
    0, \quad \textrm{for another half of pairs }(\rho,\sigma). \\
\end{array}
\right.
\end{equation}
So the Pauli operator expansion of $[X,Y]$ is dominated by terms of size $ k+l-\frac{4kl}{3N}$, which implies that
\begin{equation}
\label{eq:commutator size 1}
    \mathcal{S}(k,l)=k+l-\frac{4kl}{3N},\quad k\sim O(N).
\end{equation}
Eq.~\eqref{eq:Pauli commutator size 1} implies half of terms in the Pauli operator expansion of $[X,Y]$ are 0, so 
\begin{equation}
    D_{k,l}=\frac{1}{2}D_k D_l.
\end{equation}
Then the typical sectional curvatures are given by
\begin{equation}
\label{eq:tp sc cur 3}
   K(k,l)=\frac{-3f(k+l-\frac{4kl}{3N})+2(f(k)+f(l))}{2 f(k)f(l)}.
\end{equation}

\subsubsection{$k,\,l\sim O(1)$ }
Randomly choosing $\rho$ and $\sigma$, the probability that $|\mathcal{D}_{\rho}\cap \mathcal{D}_{\sigma}|=m$ is
\begin{equation}
    \textrm{P}(m)=\frac{\binom{N}{k+l-m}\binom{k+l-m}{m}}{\binom{N}{k}\binom{N}{l}}.
\end{equation}
Compute the difference,
\begin{equation}
    \textrm{P}(m+1)-
\textrm{P}(m)=(-1+\frac{(k+l-2m)(k+l-2m-1)}{(N-k-l+m+1)(m+1)})\textrm{P}(m).
\end{equation}
In the large $N$ limit, the difference is given by
\begin{equation}
    \textrm{P}(m+1)-
\textrm{P}(m)=(-1+O(\frac{1}{N}))\textrm{P}(m).
\end{equation}
We can rewrite it as
\begin{equation}
    \textrm{P}(m+1)\sim \frac{1}{N}\textrm{P}(m).
\end{equation}
This implies that typically $\rho$ commutes with $\sigma$. If they don't commute, then the typical size of their commutator is
\begin{equation}
    \mathcal{S}(\rho,\sigma)=k+l-1.
\end{equation}
So the Pauli operator expansion of $[X,Y]$ is dominated by terms of size $k+l-1$, which implies that
\begin{equation}
\label{eq:commutator size 2}
    \mathcal{S}(k,l)=k+l-1,\quad  k,l\sim O(1).
\end{equation}
The number of noncommutative pairs in $\Delta A^k\times \Delta A^l$ is
\begin{equation}
    D_{k,l}=\frac{2}{3}\textrm{P}(1)D_k D_l.
\end{equation}
The factor $2/3$ comes from the fact that, for a single qubit, only $2/3$ of pairs in $\Delta A^1\times \Delta A^1$ don't commute. Applying  stirling's formula $n!\sim \sqrt{2\pi n}(n/e)^n$ to P$(1)$,
\begin{equation}
    \textrm{P}(1)\sim\frac{k!\,l!}{(k+l-2)!}\frac{1}{N}.
\end{equation}
The typical sectional curvatures are then given by
\begin{equation}
\label{eq:tp cur sc 4}
    K(k,l)=\frac{2}{3N}\frac{k!\,l!}{(k+l-2)!}\frac{-3f(k+l-1)+2(f(k)+f(l))}{f(k)f(l)}.
\end{equation}
A formula similar to Eq.~\eqref{eq:tp cur sc 4} was derived for 2-local operators in \cite{Brown:2017jil}.

\end{document}